\documentclass[10pt,journal,compsoc]{IEEEtran}

\usepackage{amsmath,amsfonts}
\usepackage{algorithmic}
\usepackage{array}
\usepackage{textcomp}
\usepackage{stfloats}
\usepackage{url}
\usepackage{verbatim}
\usepackage{graphicx}
\usepackage{booktabs}
\usepackage{enumerate}
\usepackage{listings}
\usepackage{hyperref}
\usepackage{xspace}
\usepackage{balance}
\usepackage[most]{tcolorbox}
\usepackage{awesomebox}
\usepackage{multirow}
\usepackage{blindtext}
\usepackage{soul}
\usepackage{mdframed}
\usepackage[colorinlistoftodos]{todonotes}
\usepackage{caption}
\usepackage{subcaption}

\def\BibTeX{{\rm B\kern-.05em{\sc i\kern-.025em b}\kern-.08em
    T\kern-.1667em\lower.7ex\hbox{E}\kern-.15emX}}

\definecolor{keywords}{rgb}{0.5,0,0.35}
\definecolor{comments}{RGB}{0,0,113}
\definecolor{red}{RGB}{160,0,0}
\definecolor{green}{RGB}{0,150,0}
 
\newmdtheoremenv [
 outerlinewidth = 1 ,
 roundcorner = 1pt,
 leftmargin = 1,
 rightmargin = 1,
 backgroundcolor = gray!20,
 outerlinecolor = blue!70!black,
 ntheorem = false,
] {finding}{Finding}

\newtcbtheorem{obs}{Finding}{%
        theorem name,%
        colback=gray!5,%
        colframe=gray!35!black,%
        fonttitle=\bfseries,title after break={Lemma  -- \raggedleft Continued}%
    }{lem}

\newcommand{\tb}[2]{\tipbox{{\bf Finding #1}. #2}}

\newcommand{\review}[1]{\textcolor{black}{#1}}

\newcommand{\highlight}[1]{{\color{red}}#1}

\newcommand{\mas}{MAS approach\xspace}

\newcommand{\fm}[1]{\emph{#1}\xspace}

\newcommand{\gps}{\fm{gappusin}}  
\newcommand{\rmb}{\fm{revmob}}

\newcommand{\sscore}{Similarity Score\xspace}

\newcommand{\rqa}{What is the impact of considering a larger and diverse dataset on the accuracy of the \mas for malware classification?\xspace}

\newcommand{\rqc}{What is the influence of the similarity between the original and repackaged
                  versions of the apps on the performance of the \mas for malware classification?\xspace}

\newcommand{\rqd}{What is the influence of the malware family (e.g., \fm{gappusin}, \fm{kuguo}, \fm{dowgin}) on the performance of the \mas for malware classification? }

\newcommand{\repack}{RePack\xspace}
\newcommand{\amc}{AndroMalPack\xspace}

\newcommand{\appsSmall}{102\xspace}
\newcommand{\apps}{\textcolor{black}{4,076}\xspace}

\newcommand{\sds}{\texttt{SmallDS}\xspace}
\newcommand{\cds}{\texttt{LargeDS}\xspace}

\newcommand{\avt}{\texttt{avclass2} tool\xspace}
\newcommand{\vt}{\texttt{VirusTotal}\xspace}
\newcommand{\se}{security engine\xspace}
\newcommand{\ses}{security engines\xspace}

\newcommand{\fone}{F1-score\xspace}
\newcommand{\fscoreSmall}{0.89\xspace}

\newcommand{\fscore}{\textcolor{black}{0.54}\xspace}

\newcommand{\malwares}{\textcolor{black}{2,895}\xspace}
\newcommand{\malwaresP}{\textcolor{black}{71.02}}

\newcommand{\appsGps}{\textcolor{black}{1,337}\xspace}
\newcommand{\appsGpsFN}{\textcolor{black}{1,170}\xspace}

\begin{document}

\title{Scaling Up: Revisiting Mining Android Sandboxes at Scale for Malware Classification}

\author{Francisco Handrick,
        Ismael Medeiros, 
        Leandro Oliveira,
        João Calássio,
        Rodrigo Bonifácio,
        Krishna Narasimhan,
        Mira Mezini,
        Márcio Ribeiro}

\maketitle

\begin{abstract}

The widespread use of smartphones in daily life has raised concerns about privacy and security among researchers and practitioners. Privacy issues are generally highly prevalent in mobile applications, particularly targeting the Android platform---the most popular mobile operating system. For this reason, several techniques have been proposed to identify malicious behavior in Android applications, including the Mining Android Sandbox approach (\mas), which aims to identify malicious behavior in repackaged Android applications (apps).
However, previous empirical studies evaluated the \mas using a small dataset consisting of only \appsSmall pairs of original and repackaged apps.
This limitation raises questions about the external validity of their findings and whether the MAS approach can be generalized to larger datasets.
To address these concerns, this paper presents the results of a replication study focused on evaluating the performance of the \mas regarding its capabilities of correctly classifying malware from different families. Unlike previous studies, our research employs a dataset that is an order of magnitude larger, comprising \apps
pairs of apps covering a more diverse range of Android malware families. Surprisingly, our findings indicate a poor performance of the MAS approach for identifying malware, with
the \fone decreasing from \fscoreSmall for the small dataset used in the previous studies to \fscore in our more extensive dataset. Upon closer examination, we discovered that certain malware families partially account for the low accuracy of the \mas, which fails to classify a repackaged version of an app as malware correctly. Our findings highlight the limitations of the MAS approach, particularly when scaled, and underscore the importance of complementing it with other techniques to detect a
broader range of malware effectively. This opens avenues for further discussion on addressing the blind spots
that affect the accuracy of the \mas.

\end{abstract}
\section{Introduction}\label{sec:introduction}

Mobile technologies, such as smartphones and tablets, have become fundamental to how we function as a society. Almost two-thirds of the world population
uses mobile technologies~\cite{Comscore,DBLP:journals/tse/MartinSJZH17}, with the
Android Platform dominating this market and accounting for more than 70\% of the \emph{mobile
market share}, with almost 2.5 million Android applications~\footnote{In this paper, we will use the terms Android Applications, Android Apps, and Apps interchangeably, to refer to Android software applications} (apps)
available on the Google Play Store, in June 2023~\cite{Statista}.  
As popularity rises, so does the risk of potential attacks, prompting collaborative efforts from both academia and industry to design and develop new techniques for identifying malicious behavior or vulnerable code in Android apps~\cite{10.1145/3017427}.
One popular class of Android malware is based on repackaging~\cite{DBLP:conf/wcre/BaoLL18,DBLP:conf/iceccs/LeB0GL18}, where a benign
version of an app is infected with malicious code, e.g., to broadcast
sensitive information to a private server~\cite{DBLP:journals/tse/LiBK21}, and subsequently shared with users using even official app stores. 

The Mining Android Sandbox approach (\mas) was initially designed to construct sandboxes based on exploratory calls to sensitive APIs~\cite{DBLP:conf/icse/JamrozikSZ16}. The \mas operates in two distinct phases. In the first phase (exploratory phase), automated test case generation tools are utilized to abstract the behavior of an app, focusing on recording calls to sensitive APIs (Application Programming Interfaces). Subsequently, during normal app execution (exploitation phase), the generated sandbox blocks any calls to sensitive APIs that were not observed during the exploratory phase.
Prior studies~\cite{DBLP:conf/wcre/BaoLL18,DBLP:journals/jss/CostaMMSSBNR22} have investigated the effectiveness of the MAS approach in detecting potential malicious behavior in repackaged apps. These studies have also conducted comparisons of the approach's performance by employing different test case generation tools during the exploratory phase, including Monkey~\cite{Monkey}, DroidBot~\cite{DBLP:conf/icse/LiYGC17}, and Droidmate~\cite{DBLP:conf/kbse/BorgesHZ18}---bringing evidence that DroidBot outperforms other test generation tools, uncovering many potential malicious behaviors.

Nonetheless, these previous studies have two main limitations.
First, they use a small dataset of malware comprising only \appsSmall pairs of original/repackaged versions of an app---which might compromise external validity. Second, their assessments do not investigate
the impact of relevant features of the repackaged apps on the accuracy of the \mas for malware classification, including
(a) whether or not the repackaged version is a malware, (b) the similarity between the original and the repackaged versions of an app,
and (c) the malware family~\footnote{Malware families (such as \fm{gappusin}, \fm{kuguo}, \fm{dowgin}, etc.) are often used to classify 
malware in groups that share similar codebases, attack methods, and objectives~\cite{DBLP:conf/ndss/ArpSHGR14}.} when the repackaged
version of an app is a malware. These limitations compromise a broader understanding of the \mas performance. We present more details about the \mas and related work in Section~\ref{sec:background}.

To better understand the impact of these issues on previously published results, this paper presents a replication of the study conducted by Bao et al.~\cite{DBLP:conf/wcre/BaoLL18}. We aim to verify the original study's findings by executing the test case generation tool DroidBot~\cite{DBLP:conf/icse/LiYGC17} in the same settings as the original research. Unlike the original study, here we use a curated dataset of app pairs (original/repackaged versions) significantly larger than the dataset used in Bao et al.'s study. Our new dataset contains \apps pairs of original and repackaged apps. We present more details about the datasets, data collection, and analysis procedures in Section~\ref{sec:experimentalSetup}.

{\bf Negative result.} Our study reveals a significantly lower
accuracy (\fone of \fscore) of the \mas in comparison to what the \mas approach performs in the small dataset (\fone of \fscoreSmall). 
Since an accuracy of \fscore is unsatisfactory for a trustworthy malware classification technique, we conduct a set of experiments 
to understand the reasons for the lower accuracy in our dataset.
Our further assessments reveal that the \mas fails to correctly classify most samples from a specific set of malware families, particularly those from the \gps family (a particular adware class frequently appearing in repackaged apps). Out of the total of \appsGps samples within this family in our large dataset, the \mas failed to classify \appsGpsFN samples as malware correctly. Accordingly, these families are responsible for substantially reducing the recall of the \mas. We detail the results of our experiments in Section~\ref{sec:results}. 
We also discuss the implications and possible threats to the validity of our study in Section~\ref{sec:discussion} and present some final remarks in Section~\ref{sec:conclusions}. 
The main artifacts we produced during this research are available in the paper repository.

\begin{small}
  \begin{center}
    \url{https://github.com/droidxp/paper-ecoop-results}
  \end{center}
\end{small}

\section{Background and Related Work}\label{sec:background}

There are many tools available that help developers reverse engineer Android bytecode~\cite{DBLP:conf/issta/WangGMC15}.
For this reason, software developers can easily decompile trustworthy apps, modify their contents by inserting malicious code,
repackage them with malicious payloads, and re-publish them in app stores, including official ones like the Google Play Store.
It is well-known that repackaged Android apps can leverage the popularity of real apps to increase their propagation and spread malware~\cite{DBLP:journals/tse/LiBK21}. As an example, in 2016 a repackaged version of the famous Pokémon Go app was discovered less than 72 hours after the game was officially released in the United States, Australia, and New Zealand~\cite{DBLP:journals/tse/LiBK21}. The repackaged version, originated from an unofficial app store, gained full control over the victim's phone, obtaining access to main functions such as the phonebook, audio recorder, and camera.

Repackaging has been raised as a noteworthy security concern in Android ecosystem by stakeholders in the app development industry and researchers~\cite{DBLP:journals/ese/KhanmohammadiEH19}. Indeed, there are reports claiming that about 25\% of Google Play Store app content correspond to repackaged apps~\cite{DBLP:conf/sigmetrics/ViennotGN14}. Nevertheless, all the workload to detect and remove malware from markets by the stores (official and non-official ones), have not been accurate enough to address the problem. As a result, repackaged Android apps threaten security and privacy of unsuspicious Android app users, beyond compromising the copyright of the original developers~\cite{DBLP:journals/access/KimLCP19}. Aiming at
mitigating the threat of malicious code injection in repackaged apps,
several techniques based on both static and dynamic analysis of Android apps have been proposed,
including the \mas for malware classification~\cite{DBLP:conf/icse/JamrozikSZ16,DBLP:conf/wcre/BaoLL18}.

\subsection{Mining Android Sandboxes}\label{sec:android-sandbox}

A \emph{sandbox}
is a well-known mechanism to secure a system and forbid a software component from accessing
resources without appropriate permissions. Sandboxes have also been used to build an isolated
environment within which applications cannot affect other programs, the network, or other device
data~\cite{DBLP:journals/peerj-cs/MaassSCS16}. The idea of using sandboxes emerged from the
need to test unsafe software, possible malware, without worrying about the integrity of the
device under test~\cite{DBLP:conf/esorics/BordoniCS17}, shielding the operating system from security issues.
To this end, a sandbox environment should have the minimum requirements to run the
program, and make sure it will never assign the program more privileges than it should have,
respecting the \emph{least privilege} principle. Within the Android ecosystem, sandbox approaches ensure the principle
of the \emph{least privilege} by preventing apps from having direct access to resources like device hardware (e.g., GPS, Camera), or sensitive data from other apps. Access to sensitives data
like contact list or resources are granted through specific APIs, known as sensitive APIs, which are managed by coarse-grained Android permissions system~\cite{DBLP:journals/corr/abs-2109-06613}.

The \mas~\cite{DBLP:conf/icse/JamrozikSZ16} aims at automatically
building a sandbox through dynamic analysis (i.e., using automatic test generation tools).
The main idea is to grant permissions to an app based on its calls to sensitive APIs.
Thus, sandboxes build upon these calls to create safety rules and then block future
calls to other sensitive resources, which diverge from those found in the first exploratory
phase. Using the Droidmate test generation tool~\cite{DBLP:conf/icse/JamrozikZ16},
Jamrozik et al. proposed a full-fledged
implementation of the \mas, named Boxmate~\cite{DBLP:conf/icse/JamrozikSZ16}. 
Boxmate records the occurrences of calls to sensitive APIs and, optionally, the UI
events (e.g., a button click) that trigger these calls. Therefore, it is possible to configure Boxmate to record events associated with each sensitive call as
tuples (event, API), instead of recording just the set of calls to sensitive APIs. Jamrozik et al. argue that, in this way, Boxmate generates finer
grain results, which
might improve the accuracy of the \mas---even with the presence of reflection, a feature commonly used in
malicious apps~\cite{DBLP:conf/issta/0029BOK16}.

In fact, the \mas can be implemented using
a mix of static and dynamic analysis. In the first phase, one
can instrument an Android app to log any call to the Android sensitive methods.
After that, one can execute a test case generation tool (such as DroidMate, DroidBot, or Monkey) to explore the app behavior at runtime,
while the calls to sensitive APIs are recorded. This set of calls to sensitive APIs is then used
to configure the sandbox. The general \mas suggests that the more efficient the test generation tool (for instance, code coverage),
the more accurate the resulting sandbox would be.

\subsection{Mining Android Sandbox for Malware Classification}\label{sec:classification}

Besides being used to generate Android sandboxes, the \mas can also be used  
to detect if a repackaged version of an Android app contains an unexpected (perhaps malicious)
behavior~\cite{DBLP:conf/wcre/BaoLL18}. In this scenario, the \emph{effectiveness} of the approach
is estimated in terms of the accuracy in which malicious behavior is correctly identified in the repackaged version of the
apps.

The \mas for malware classification typically works as follows. In a first step ({\bf instrumentation phase}), 
a tool instruments the code of the apps (both original and repackaged versions) to collect relevant information
during the apps execution in later stages. Then, in a second step ({\bf exploration phase}),
the \mas collects a set $S_1$ with all calls to sensitive APIs the original version of an app executes while running a test case generator tool (like DroidBot).
In the third step ({\bf exploitation phase}), the \mas (a) collects a set $S_2$ with all calls to sensitive APIs the repackaged version of an app
executes while running a test case generator tool, and then (b) computes the set $S = S_2 \setminus S_1$ and checks whether  $S$ is empty or not.
The \mas classifies the repackaged version as a malware whenever $|S| > 0$.  

Previous research works reported the results of empirical studies that aim to investigate the effectiveness of
the \mas for malware classification~\cite{DBLP:conf/wcre/BaoLL18,DBLP:conf/scam/CostaMCMVBC20}.
For instance, Bao et al. found that, in general, the sandboxes constructed using test generation tools classify at least 66\% of repackaged apps as malware in a
dataset comprising 102 pairs of apps (original/repackaged versions)~\cite{DBLP:conf/wcre/BaoLL18}.
Actually, the mentioned work performed two studies: one pilot study involving a dataset
of 10 pairs of apps (\texttt{SmallE}), in which the authors executed each test case generation tools for one hour; and a larger experiment
(\texttt{LargeE}) involving 102 pairs of
apps in which the authors executed the test case generation tools for one minute~\cite{DBLP:conf/wcre/BaoLL18}.

The authors also presented that, among five test generation tools used, DroidBot~\cite{DBLP:conf/icse/LiYGC17} leads to the most effective sandbox.
Le et al. extend the \mas for malware classification with additional verification,
such as the values of the actual parameters used in the
calls to sensitive APIs~\cite{DBLP:conf/iceccs/LeB0GL18}, while
Costa et al.\cite{DBLP:journals/jss/CostaMMSSBNR22} investigated the impact of static analysis to complement the accuracy of the \mas
for malware classification. Their study reports that DroidFax~\cite{DBLP:conf/icsm/CaiR17a}, the static analysis infrastructure used in~\cite{DBLP:conf/wcre/BaoLL18}, classifies as malware almost half of the repackaged apps.

\subsection{Android Malware Classification}

The field of malware detection for the Android platform is fertile, with a significant number of secondary studies already published~\cite{DBLP:conf/eann/SerajPP22,DBLP:conf/icsoft/LekssaysFA20,DBLP:journals/access/WeiLWZZY17,DBLP:conf/codaspy/ZhouZJN12}. In general, malware detection techniques are divided into static detection, dynamic detection, and hybrid detection~\cite{choudhary2018haamd}. Several studies have also conducted surveys on malware detection techniques and presented a review of them~\cite{DBLP:journals/access/LiuXXZSL20,DBLP:journals/csur/TamFASC17,DBLP:conf/icai2/OdusamiAMSDM18}. For instance, M. Odusami et al.~\cite{DBLP:conf/icai2/OdusamiAMSDM18} discuss various static analyses approaches that have been used in the literature to identify malicious behavior in Android apps. The authors present some works with permission and signature-based malware detection systems. They highlight that both approaches have a low false positive rate; however, they are very ineffective in detecting new malware. Although they could reveal possible malicious behaviors, the authors discuss several limitations of these approaches, as they are limited regarding code obfuscation and dynamic code loading.

The literature also presents surveys based on dynamic analysis, where malicious behavior is analyzed at runtime, exposing risks that are not detected by static analysis. As a malicious app ``is alive'', dynamic analysis adds another degree of analysis since it observes how Android apps interacts with the environment. However, if applied inappropriately, it may provide limited code coverage, which repeated executions can improve. Therefore, dynamic analysis's time cost and computation resources are higher when compared with static analysis. K. Tam et al.~\cite{DBLP:journals/csur/TamFASC17} presented several dynamic analysis studies based on Android architectural layers. The survey also exposed that dynamic analysis can be performed in emulator environments, real devices, or both. The authors discuss that the choice of environments is an important issue for analysis, as there are malware families that can detect emulated environments and do not exhibit malicious behaviors~\cite{DBLP:journals/csr/SihagVS21}. Finally, K. Tam et al. also exposed some works based on hybrid malware detectors and claim that these methods can increase code coverage and robustness, taking advantage of the best of each technique to find malicious behaviors.

Several studies have also explored Android malware detection approaches based on machine learning (ML) techniques, employing both static and dynamic analyses to extract features and train ML models~\cite{DBLP:journals/access/LiuXXZSL20}. Most of these approaches have demonstrated high accuracy (above 90\%), effectively detecting previously unseen malware families with low false positives rate~\cite{DBLP:journals/corr/abs-2001-09406}. However, some studies have identified limitations of machine learning approaches for Android malware classification. In their work, K. Liu et al.~\cite{DBLP:journals/access/LiuXXZSL20} highlighted challenges related to machine learning techniques, identifying several factors that could lead to biased results, such as the quality of the sample set. The authors argue that samples of poor quality, with a non-representative size or outdated samples, may yield promising results in experimental settings but might not perform similarly in a real environment~\cite{DBLP:journals/access/LiuXXZSL20}. Another critical aspect is the quality of the extracted feature dataset. The efficacy of machine learning approaches heavily relies on the selection of correct features and their extraction methods, particularly dynamic features. Moreover, in addition to the computational costs involved, other studies~\cite{DBLP:journals/tdsc/DemontisMBMARCG19,DBLP:journals/csur/GardinerN16} have indicated that machine learning approaches exhibit weaknesses when dealing with malicious apps that alter their behavior to mislead learning algorithms, which may restricts their applicability in real-world scenarios.
\section{Experimental Setup}\label{sec:experimentalSetup}

This research aims to develop a deeper understanding of
the performance of the \mas for detecting malware. To this
end, in this paper, we {\bf replicate} the study by Bao et al.~\cite{DBLP:conf/wcre/BaoLL18}, which advocates for using the \mas for malware classification. However, in contrast to the original study~\cite{DBLP:conf/wcre/BaoLL18}, we use a dataset of repackaged apps that is an order of magnitude larger.
Accordingly, we investigate the following research questions:

\begin{enumerate}[(RQ1)]
\item \rqa Answering this question may help shed light on potential generalization issues in previous studies that empirically assess the \mas approach to malware classification.

\item \rqc Answering this research question helps clarify whether the similarity between an original app and its repackaged version affects the \mas approach's performance in malware classification.  

\item \rqd Answering this research question may help identify potential blind spots in the \mas approach to malware classification, revealing possible extensions that could improve the detection of specific malware families.

\end{enumerate}

In this section, we describe our study settings. First, we present our procedures to create our datasets (Section~\ref{sec:dataset}).  Then, we describe the data collection and data analysis procedures (Sections~\ref{sec:dataCollectionProc} and~\ref{sec:dataAnalysisProc}).

\subsection{Malware Dataset}\label{sec:dataset}

To address our research questions, we contribute a dataset designed to meet two primary requirements.
First, it should provide a comprehensive and up-to-date selection of Android repackaged apps. By ``comprehensive'', we mean at least an order of magnitude larger than the dataset used in the original study~\cite{DBLP:conf/wcre/BaoLL18}. Given its comprehensiveness, we expect it to include a diverse range of malware families to ensure representativeness.
Second, our dataset should be properly labeled, ensuring each sample includes key attributes such as similarity and malware family. This is particularly necessary to answer research questions RQ2 and RQ3. 

\subsubsection{Procedures for Building the Dataset}

We curate our dataset in three main phases. 
In the first phase, we use two repositories of repackaged Android apps (\repack~\cite{DBLP:journals/tse/LiBK21} and \amc~\cite{rafiq2022andromalpack}) to build the
dataset we use in our research. \repack was curated using automatic procedures that extract repackaged apps from the Androzoo
repository~\cite{DBLP:conf/msr/AllixBKT16}. It comprises 18,073 apps, from which 2,776 are original versions of an app and the remaining ones are repackaged. \repack contains 15,297 pairs of original and repackaged Android
apps, many repackaged versions of the same original app may coexist within the \repack dataset---note that all repackaged variants of a given app are derived from the same original version, as confirmed by their matching hash identifier. \repack is the leading dataset used in Android
repackaged research~\cite{DBLP:journals/ese/KhanmohammadiEH19}, even though it only contains packages built until 2018. For this reason, we decided to include samples from the \amc dataset collected after 2018 in our research. Unlike RePack, \amc lacks information about the original apps, leading
us to follow an existing heuristic~\cite{DBLP:journals/tse/LiBK21} to identify the original versions of its repackaged apps,
leading to a sample from the \amc dataset that contains 1,190 pairs (original/repackaged) of apps, all pairs
satisfying our constraint of being collected after 2018. Altogether, our initial dataset contains a total of 16,487 pairs of apps.

In the second phase, we discarded some samples from our initial dataset because, during the execution of our experiments, we encountered recurrent issues related to the instrumentation of the apps using DroidFax~\cite{DBLP:conf/icsm/CaiR17a}.
Other problems occurred after the execution of the apps in the Android emulator, while analyzing the apps or their execution logs.
More precisely, we encountered failures while instrumenting 919 original apps from our initial dataset, including both \repack and \amc. After removing these original apps
from our dataset, we were left with 5,875 pairs (original/repackaged) of apps. Among these pairs, 430 repackaged apps could not be instrumented.
Failures also occurred while analyzing either the original or repackaged version of 586 apps, resulting in a dataset containing 4,742 pairs. Failures at this phase were expected, as some malware samples employ evasion tactics, such as deliberately crashing test apps in simulated environments, to avoid detection~\cite {DBLP:journals/corr/abs-2403-16304}. Finally, we could not install five apps in the version of the Android emulator (API level 28) we used in our research.
Compared to our experience building our dataset, a more significant percentage of failures has been reported in previous research~\cite{DBLP:conf/wcre/BaoLL18}. Note that we did not apply any filters to increase the representation of certain malware families in our dataset.

Third, we queried the \vt repository
to identify original versions of apps labeled as malware. Samples with such labels were excluded from our dataset,
as the \mas assumes that the original version of an app is not malware (otherwise, the repackaged versions might
also exhibit malicious behavior). \vt is a widely recognized tool that scans software assets, including Android apps,
using over 60 antivirus engines~\cite{DBLP:journals/ese/KhanmohammadiEH19}. Thus, we excluded 661 samples from our dataset that
do not satisfy this constraint.

In the end, we are left with our final dataset (hereafter \cds) of \apps apps which we use in our study. 
To bring evidence that we were able to reproduce the results of previous research, we also consider in our research
a small dataset (\sds) used in the original study~\cite{DBLP:conf/wcre/BaoLL18}.
This is the same dataset referenced in Section~\ref{sec:classification} as (\texttt{LargeE}).

\subsubsection{Features of the Datasets}
 
We queried the \vt repository to find out which repackaged apps in our
dataset have indeed been labeled as a malware. According to \vt, in the \sds (102 pairs),  
69 of the repackaged apps (67.64\%) were identified as malware by at least two \ses.  
Here, we consider a repackaged version of an app to be malware only if \vt reports that at least two \ses identify malicious behavior within the asset.  
Although this decision aligns with previous research~\cite{DBLP:conf/uss/ZhuSYQZS020,DBLP:journals/ese/KhanmohammadiEH19},  
we assess its potential impact on our findings in Section~\ref{sec:discussion}.
Considering the \cds, at least two security engines identified \malwares out of the \apps repackaged apps as malware (\malwaresP\%). 
Again, in Section~\ref{sec:discussion}, we show that our results remain consistent across three additional scenarios:  
classifying a repackaged version of an app as malware if at least one, five, or ten \vt engines flag it as malicious.

Classifying malware into different categories is a common practice. For instance, Android malware can be classified into categories
like riskware, trojan, adware, etc. Each category might be further specialized in several malware families, depending on its
characteristics and attack strategy---e.g., steal network info (IP, DNS, WiFi), collect phone info,
collect user contacts, send/receive SMS, and so on~\cite{DBLP:conf/iccns/RahaliLKTGM20}.
According to the
\avt~\cite{DBLP:conf/acsac/SebastianC20}, the malware samples in the \sds come from 17 different families---most of them from the Kuguo (49.27\%) and Dowgin (17.39\%) families.  
Our \cds, besides comprising a large sample of repackaged apps (\apps in total),
contains \review{116} malware families---most
of them from the Gappusin (46.18\%) family. Despite being flagged as malicious by at least two security engines, unfortunately \avt cannot correctly identify the family of 253 samples in our \cds.

We also characterize our dataset according to the similarity
between the original and repackaged versions of the apps, using the  
SimiDroid tool~\cite{DBLP:conf/trustcom/0029BK17}. SimiDroid quantifies the similarity
based on (a) the methods that are either identical or similar in both versions of the apps (original and repackaged versions),
(b) methods that only appear in the repackaged version of the apps (new methods), and (c) methods that only appear in the
original version of the apps (deleted methods).
Our \cds has an average similarity score of 90.39\%, with the following distribution:  
87 app pairs have a similarity score below 25\%, 49 pairs fall between 25\% and 50\%,  
353 apps between 50\% and 75\%, and 3,587 apps exceed 75\%.  
The \sds has an average similarity score of 89.41\%.  

After executing our experiments, we identified the most frequently abused sensitive APIs called by the repackaged version of our samples.
We observed that upon execution of all samples from our dataset (\sds and \cds), malicious app versions injected \review{133} distinct methods from sensitive APIs (according to the
AppGuard~\cite{DBLP:conf/esorics/BackesGHMS13} security framework).
Malicious code often exploits these APIs to compromise system security and access sensitive data.  
Table~\ref{tab:APIused} lists the 10 most frequently called methods from sensitive APIs that appear only in the repackaged versions of the apps.

\renewcommand{\arraystretch}{1.2}
\begin{table*}[htb]

\centering
\begin{small}

\begin{tabular}{lp{1cm}l}\toprule

    Method of Sensitive API & Occurrences \\ \midrule
    
    android.telephony.TelephonyManager: int getPhoneType() &  311\\
    android.telephony.TelephonyManager: java.lang.String getNetworkOperatorName() &  297 \\
    android.location.LocationManager: java.lang.String getBestProvider(android.location.Criteria,boolean) &  292 \\
    android.telephony.TelephonyManager: int getSimState() &	284\\
 java.lang.reflect.Field: java.lang.Object get(java.lang.Object)&	277\\
    android.net.NetworkInfo: java.lang.String getTypeName() &  271\\
    android.database.sqlite.SQLiteDatabase: android.database.Cursor query(java.lang.String,java.lang.String[],...,...,...,...,...) &  270 \\
       java.lang.reflect.Field: int getInt(java.lang.Object) &  250\\
        android.net.wifi.WifiInfo: java.lang.String getMacAddress()	& 238\\
    
    android.telephony.TelephonyManager: java.lang.String getNetworkOperator() &  237\\

\bottomrule 

\end{tabular}

\end{small}
\caption{Sensitive APIs that frequently appear in the repackaged versions of the apps.}\label{tab:APIused}
\end{table*}

\renewcommand{\arraystretch}{1}

We must highlight that the \cds samples come from different Android app stores. Most of our repackaged apps come from a non-official
Android app store, Anzhi~\cite{anzhi}. However, some repackaged apps also come from the official Android app store, Google Play.

\subsection{Data Collection Procedures} \label{sec:dataCollectionProc}

We take advantage of the DroidXP infrastructure~\cite{DBLP:conf/scam/CostaMCMVBC20}
for data collection. DroidXP allows researchers to compare 
test case generation tools for malicious app behavior identification, using the \mas. Although the comparison of test
case generation tools is not the goal of this paper, DroidXP
was still useful for automating the following steps of our study.

\begin{enumerate}[(Step1)]
 \item \textbf{Instrumentation}: In the first step,
we configure DroidXP to instrument all pairs of apps in our datasets (\sds and \cds).
Here, we instrument both versions of the apps (as APK files) to collect relevant information during their execution. Under the hood, DroidXP leverages
DroidFax to instrument the apps and collect static
information about them. To improve the performance across multiple executions,
this phase executes only once for each version of the apps in our dataset.

\item \textbf{Execution}: In this step, DroidXP first installs the (instrumented) version of the APK files in the Android emulator we use in our experiment (API 28) and then starts a test case generation tool for executing both app versions (original and repackaged). We execute the apps via DroidBot~\cite{DBLP:conf/icse/LiYGC17}, mainly because the original research we replicate here reports that DroidBot leads to the best accuracy of the \mas for malware identification. Since previous studies suggest that DroidBot's coverage nearly reaches its maximum within one minute~\cite{DBLP:conf/wcre/BaoLL18}, we run each app for three minutes. To mitigate the randomness inherent in test case generation tools, we repeat this process three times. To also ensure that each execution gets the benefit of running on a fresh Android instance without biases that could stem out of history, DroidXP wipes out all data stored on the emulator that has been collected from previous executions.

\item \textbf{Data Collection}: After the execution of the instrumented apps, once again, DroidXP leverages DroidFax, this time to collect all relevant information (such as calls to sensitive APIs, test coverage metrics, and so on). We use this information to analyze the performance of the \mas for detecting malicious behavior.
\end{enumerate}

\subsection{Data Analysis Procedures} \label{sec:dataAnalysisProc}

We consider that the \mas builds a sandbox that labels a repackaged version
of an app as malware if there is at least one call to a sensitive API that (a) was observed
while executing the repackaged version of the app and that (b) was not observed while
executing the original version of the same app. If the set of sensitive methods that only the repackaged version of an app calls is empty,
we conclude that the sandbox does not label the repackaged version of an app as malware. The set of sensitive APIs we use was defined in the AppGuard framework~\cite{DBLP:conf/esorics/BackesGHMS13}, which was based on the mapping from sensitive APIs to permissions proposed by Song et al.~\cite{DBLP:conf/ccs/FeltCHSW11}. We triangulate
the results of the \mas classification with the outputs of \vt, which might lead to one of the following
situations:

\begin{itemize}
\item {\bf True Positive (TP)}. The \mas labels a repackaged version as malware and, according to
  \vt, at least two \ses label the asset as a malware.
  
\item {\bf True Negative (TN)}. The \mas does not label a repackaged version as malware and,
  according to \vt, at most one \se labels the asset as a malware. 

\item {\bf False Positive (FP)}. The \mas labels a repackaged version as malware and, according to
  \vt, at most one \se labels the asset as a malware.

\item {\bf False Negative (FN)}. The \mas does not label a repackaged version as malware, and
  according to \vt, at least two \ses label the asset as a malware.
\end{itemize}

In Section~\ref{sec:results} we compute \emph{Precision}, \emph{Recall}, and \emph{F-measure} ($F_1$) from
the number of true-positives, false-positives, and false-negatives (using standard
formulae). We use basic statistics (average, median, standard deviation) to identify the
accuracy of the \mas for malware classification, using both datasets---i.e., the \sds
with 102 pairs of apps and \cds with
\apps pairs. We use the Spearman Correlation~\cite{spearman-correlation} method and
Logistic Regression~\cite{statistical-learning} to understand the strengths of
the associations between the similarity index between the original and the repackaged versions
of a malware with the \mas accuracy---that is,
if the approach was able to classify an asset as malware correctly. We also use existing tools to reverse engineer a sample of repackaged
apps in order to better understand the (lack of) accuracy
of the \mas. 

Table~\ref{tab:replication} highlights the differences between the original study~\cite{DBLP:conf/wcre/BaoLL18} and our replication study. In the best-case scenario, where no re-executions are required, our experiment would take at least 611 hours. In contrast, under the best conditions, the original experiment's execution would last 60 hours. This difference is one of the reasons we focus our research on DroidBot, the test case generation tool that demonstrated the best performance in the original study.

\renewcommand{\arraystretch}{1.2}
\begin{table*}[htb]
\centering
\begin{small}
\begin{tabular}{lp{4cm}l}\toprule
{\bf Study Feature} & {\bf Original Study} & {\bf Replication Study} \\ \midrule
Dataset & 102 pairs of samples & 4,076 pairs of samples \\ 
Execution time & One minute & Three minutes \\ 
Number of executions & Single execution & Three executions \\ 
Metrics & Malware prevalence & Precision, Recall, and \fone \\ 
Test case generation tool & Six different tools 
                            (DroidBot with best performance) & DroidBot only \\
\bottomrule 
\end{tabular}
\end{small}
\caption{Characterization of this replication study}
\label{tab:replication}
\end{table*}

\renewcommand{\arraystretch}{1}

\section{Results}\label{sec:results}

In this section, we detail the findings of our study.  We remind the reader that this replication study's primary goal is to better understand the strengths and limitations of the \mas for malware detection by replicating the work of Bao et al., using DroidBot as the test case generation tool. We explore
the results of our research using two datasets: the \sds (\appsSmall app pairs), and \cds (\apps pairs).

\subsection{Exploratory Data Analysis of Accuracy}\label{sec:accuracy}

{\bf \sds.} Considering the \sds (102 apps), the \mas for malware detection 
classifies a total of 69 repackaged versions as malware (67.64\%).
This result is close to what Bao et al. reported~\cite{DBLP:conf/wcre/BaoLL18}.
That is, in their original paper,  the \mas using DroidBot classifies 66.66\% of the
repackaged version of the apps as malware~\cite{DBLP:conf/wcre/BaoLL18}.
This result confirms that we could reproduce the findings of the original study using our implementation settings of the \mas. 

\tb{1}{We were able to reproduce the results of
  existing research using our implementation of the \mas,
  achieving a malware classification in the
  \sds close to what has been reported in
  previous studies.}

In the original study~\cite{DBLP:conf/wcre/BaoLL18},
the authors assume that all repackaged versions are malware and contain 
a malicious code. For this reason, the authors do not
explore accuracy metrics (such as Precision, Recall, and
F-measure ($F_1$))---all repackaged apps labeled as
malware are considered true positives in the original study.
As we mentioned, in this paper we take advantage
of \vt to label our dataset and build a ground truth:
In our datasets, we classify a repackaged version of an app as malware if, according to our \vt
query results, at least two security engines identify malicious behavior in the asset.
This decision aligns with existing recommendations~\cite{DBLP:conf/uss/ZhuSYQZS020,DBLP:journals/ese/KhanmohammadiEH19}). The first row of Table~\ref{tab:accuracy} shows that the \mas achieves an accuracy of \fscoreSmall when
considering the \sds. Nonetheless, the \mas fails
to classify seven assets as malware on the \sds correctly (FN column, first row of Table~\ref{tab:accuracy}),
and wrongly labeled the repackaged version of six apps as malware (FP column).

\begin{table*}[htb]
  \caption{Accuracy of the \mas in both datasets.}
\centering{
  \begin{tabular}{lrrrrrr} \hline
    Dataset & TP   & FP  & FN  & Precision & Recall & $F_1$ \\
    \hline
    \sds (102)    & 63   & 6   & 7   & 0.91      & 0.90   & 0.90  \\
    \cds (4,076)    & 1,175  & 220 & 1,720 & 0.84       & 0.40   & 0.54  \\
    \hline
  \end{tabular}
  }
  \label{tab:accuracy}
\end{table*}

{\bf \cds.} Surprisingly, considering our complete dataset (\apps apps), the \mas
labels a total of \review{1,395} repackaged apps as malware (\review{34.22\%} of the total number of repackaged
apps)---for which the repackaged version calls at least one additional sensitive API.
Our analysis also reveals a {\bf negative result} related to the accuracy of the approach: here,
the accuracy is much lower in comparison to what we reported for the
\sds (see the second row of Table~\ref{tab:accuracy}): $F_1$ dropping from \fscoreSmall to \fscore.
This result indicates that, when considering a large dataset, the accuracy of the \mas using
DroidBot drops significantly.

\tb{2}{
  The \mas for malware detection
  leads to a substantially lower performance on the
  \cds (\apps pairs of apps),
  dropping \fone from \fscoreSmall to \fscore in comparison to
  what we observed in the \sds.}

Therefore, the resulting sandbox we generate using
DroidBot suffers from a significantly low accuracy rate when considering a large dataset. This is shown in the second row of Table~\ref{tab:accuracy}.
The negative performance of the \mas in the \cds encouraged us to endorse efforts to identify potential reasons for
this phenomenon and motivated us to explore the research questions RQ2 and RQ3.

\subsection{Assessment Based on \sscore}

\begin{figure}[t!]
  \includegraphics[width=\columnwidth]{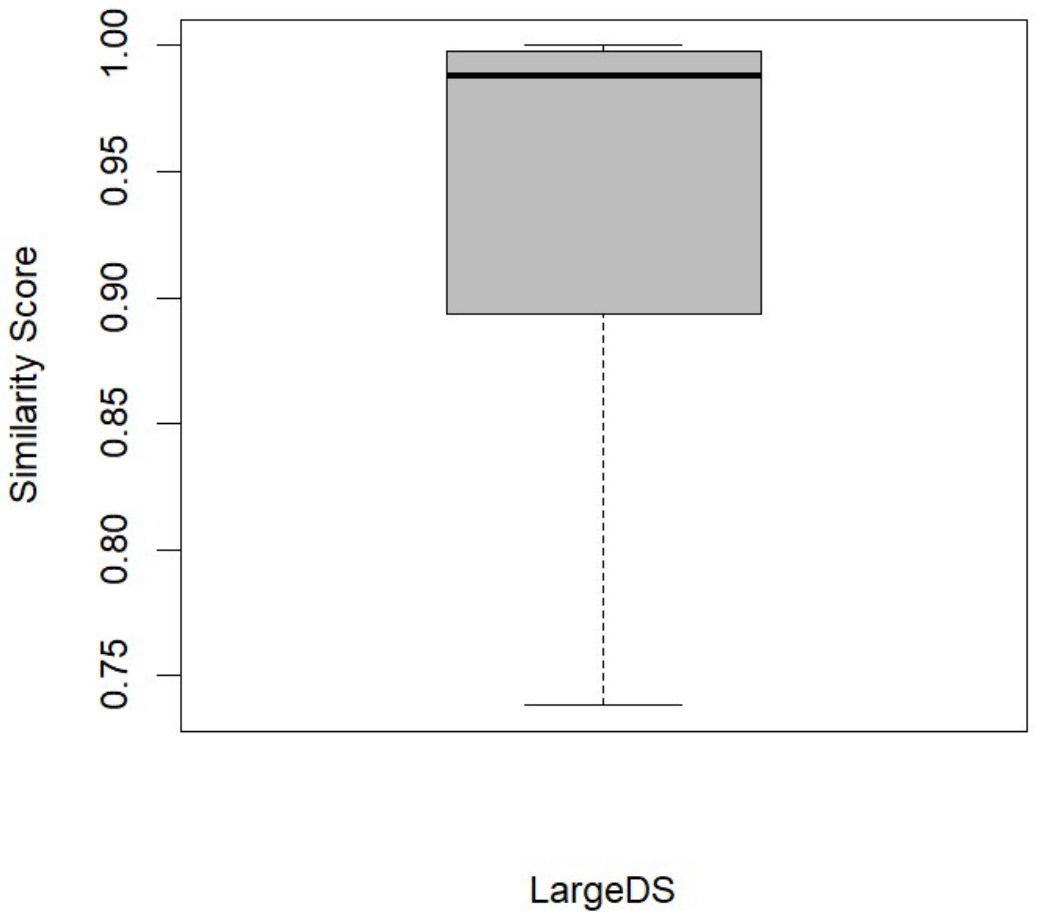}
  \caption{\sscore of the malware samples in the \cds. The boxplots in the figure do not show
  outliers.}
  \label{fig:ss}
\end{figure}

Figure~\ref{fig:ss} shows the \sscore distribution
over the \cds we use in our research.
Recall that the \sscore measures how similar an app's
original and repackaged versions are.
The complete dataset averages a \sscore of 0.90 (with a median of 0.98 and standard deviation of 0.18). 

In this section we investigate how the \sscore influences the accuracy of the \mas---which might help us understand if it might explain the low small performance of the \mas in the \cds. To this end, we leverage
Logistic Regression to quantify the relationship between \sscore and \fone. This analysis excludes instances of true
negatives (i.e., cases where the repackaged version is benign according to \vt and the \mas
correctly labels it as benign).
As such, we test the following null hypothesis:

\begin{quote}
  {\bf $H_0$} \sscore does not influence the accuracy of the
  \mas for malware detection. 
\end{quote}

The logistic regression results suggest that we should reject our null hypothesis ($p$-value $=$ $2.22\cdot10^{-16}$).
This finding indicates that the accuracy of the \mas on \cds is influenced by the similarity between the original and repackaged versions
of an app.

\tb{3}{There is an association between
  the \sscore and the \mas performance, which means that
  the similarity between the original and repackaged versions of
  an app can explain the performance
  of the \mas for malware classification.}

To clarify the association between
\sscore and accuracy, we use the \emph{K-Means} algorithm to split the
\cds into ten clusters---according to the \sscore. We then
estimate the percentage of correct classifications for each cluster, as
shown in Table~\ref{tab:ss-clusters}. Note that the \mas
achieves the highest percentage of correct classification (77.35\%) for the second cluster (cId = 2), which
presents an average \sscore of (0.56). Nonetheless, the
cluster cId = 10, with a larger number of samples (1,302) and \sscore (0.99), presents a percentage of correct classifications of 26.5\%. We can observe that as the average similarity rate decreases, there is a tendency toward greater accuracy in the \mas. Hence, the average similarity score could explain the poor performance of the \mas on the \cds, especially considering that most samples exhibit a high average similarity of 0.99.

\begin{table}[ht]
  \caption{Characteristics of the clusters. Note there is a specific
    pattern associating the percentage of
    correct answers with the \sscore.
   For this analysis, we removed the true negatives in our dataset.}
  \centering
  \begin{small}
    \begin{tabular}{rrrrr}   \hline
 cId & \sscore & Samples & Correct Answers & \% \\ \hline

1 &   0.42 &  42 &  30 & 71.42 \\ 
  2 &   0.56 & 181 & 140 & 77.35 \\ 
  3 &  0.68 & 131 &  98 & 74.81 \\ 
  4 &   0.80 & 170 & 104 & 61.18 \\ 
  5 &   0.88 & 263 &  83 & 31.56 \\ 
  6 &   0.91 & 236 & 129 & 54.66 \\ 
  7 &   0.95 & 167 &  51 & 30.54 \\ 
  8 &   0.97 & 150 &  67 & 44.67 \\ 
  9 &   0.98 & 421 & 112 & 26.60 \\ 
  10 &   0.99 & 1302 & 345 & 26.50 \\
   \hline

 \end{tabular}
 \end{small}
 \label{tab:ss-clusters}
 \end{table}

\subsection{Assessment Based on Malware Family}

As we discussed in the previous
section, the similarity assessment partially explains the low performance of the
\mas on the \cds. Since the \cds covers a wide range of malware families, we investigate the
hypothesis that the diversity of malware families in
the \cds also contributes to
the poor performance of the \mas on the \cds.
Indeed, in the \cds, we identified a total of 116 malware families, though the most frequent
ones are \gps (\appsGps samples), \fm{revmob} (207 samples), \fm{dowgin} (183 samples) and \emph{airpush} (120 samples). Together, they
account for \review{63.79\%} of the repackaged apps in our \cds labeled as malware according to \vt.

This family distribution in the \cds is
different from the family
distribution in the \sds (used in the original study)---where the
families \fm{kuguo} (34 samples), \fm{dowgin} (12 samples),
and \fm{youmi} (5 samples) account for
73.91\% of the families considering the 69
repackaged apps in the \sds for which \vt labels as malware.
Most important, in the \sds, there is just one
sample from the \gps family and no sample from \fm{revmob} family, two of the most frequent families in our \cds. This observation
leads us to the question: how does the \mas
perform when considering only samples from the \gps and \fm{revmob} families?

The confusion matrix of Table~\ref{tab:gappusin} summarizes the accuracy assessment of the \mas considering
only the \gps and \fm{revmob} samples in the \cds. To make clear, \vt classifies as malware all repackaged versions in the \gps and \fm{revmob}
family. It is worth noting that the \mas failed to classify correctly
1,170 (87.5\%) samples of \gps as malware. Similarly,
92 samples (44.44\%) from \fm{revmob} were not classified as malware.
Furthermore, if we exclude the \gps and \fm{revmob} samples from the \cds,
the recall of the \mas increases to 0.72, which, although improved, remains relatively low compared to the original studies.

\begin{table}[ht]
  \caption{Confusion matrix of the \mas when considering only the
  samples from the \gps and \fm{revmob} family in the \cds.}
\centering
\begin{tabular}{r|cc} \hline
\multirow{2}{*}{Actual Condition}   & \multicolumn{2}{c}{Predicted Condition} \\ 
                                    & Benign    & Malware   \\ \hline 
  Benign   (0)                       & TN (0)    & FP (0)    \\
  Gappusin (\appsGps)                     & FN (\appsGpsFN)  & TP (167)   \\
  Revmob (207)                     & FN (92)  & TP (115)   \\
  \hline
\end{tabular}
\label{tab:gappusin}
\end{table}

\review{
  \tb{4}{The \mas fails to correctly identify 87.50\% of the samples from the \gps family and 44.44\% of
    the samples from the \fm{revmob} as malware. Just like the Similarity Score, the presence of some malware
    families with a high false negative rate also influences the low recall of the \mas in the \cds}
  }

\begin{figure}[ht]
  \centering
  \begin{tabular}{@{}c@{}}
    \includegraphics[scale=0.3]{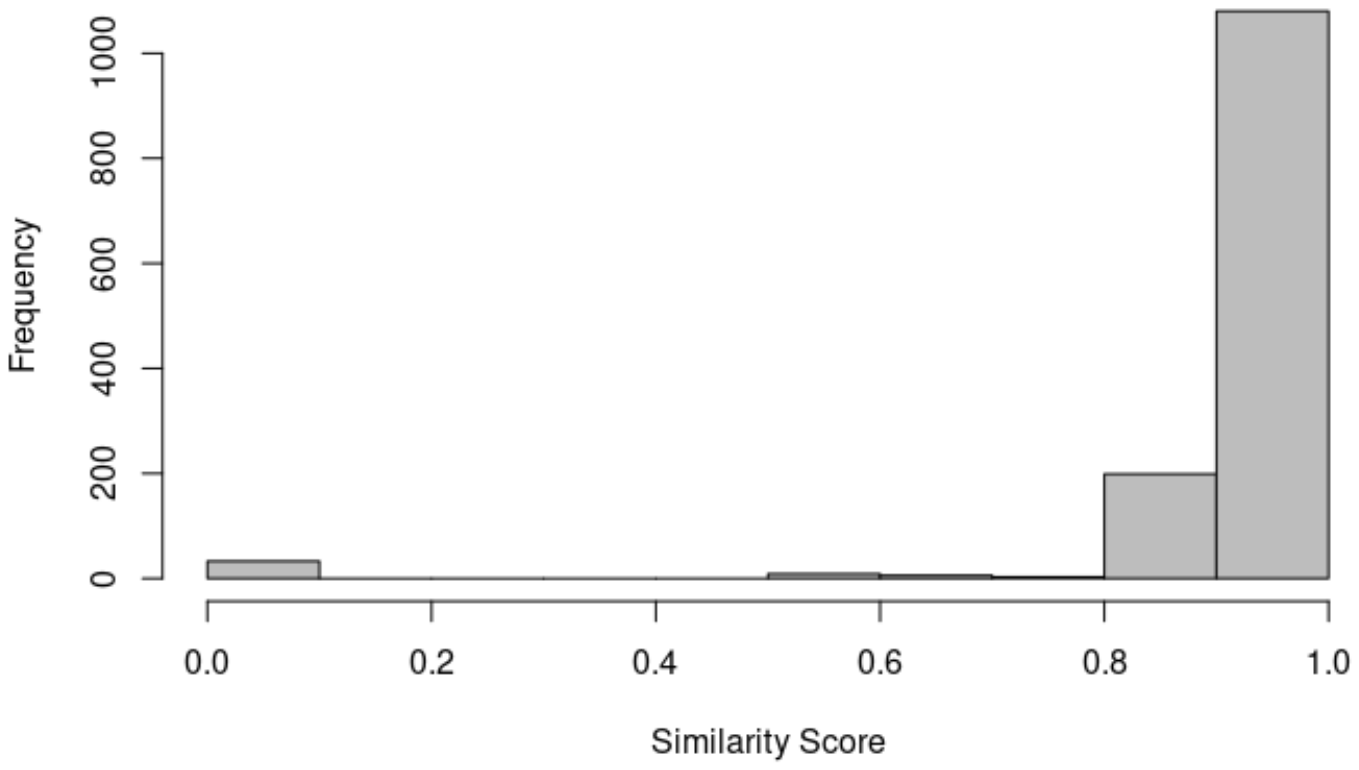} \\[\abovecaptionskip]
    \small (a) \sscore for the samples in the \gps family.
  \end{tabular}

  \begin{tabular}{@{}c@{}}
    \includegraphics[scale=0.3]{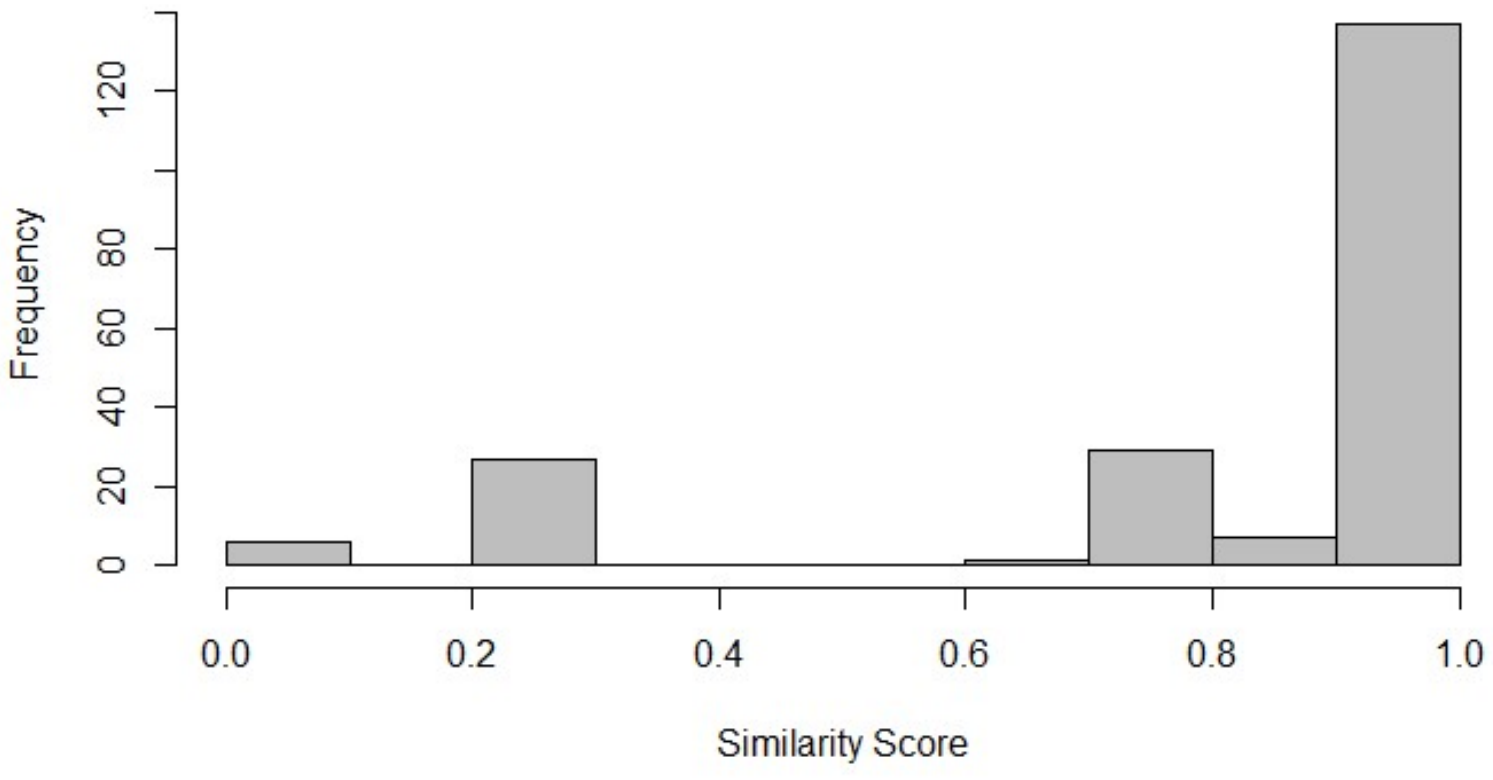} \\[\abovecaptionskip]
    \small (b) \sscore for the samples in the \rmb family.
  \end{tabular}

  \caption{Histogram of the \sscore for the samples in the \gps and \rmb families.}\label{fig:hist}
\end{figure}

We further analyze the samples from the \gps and \rmb malware families in our dataset, given their relevance to the negative results presented in our paper. First, we examined the \sscore of the samples. Figure~\ref{fig:hist} shows a histogram of the \sscore for both families. Most repackaged versions are similar to the original ones, with an average \sscore of 0.94, a median of 0.99, and a standard deviation (SD) of 0.16 for the \gps family. For the \rmb family, the average \sscore is 0.81, the median is 0.91, and the SD is 0.26.

We also reverse-engineered samples from both families. Due to the significant effort required for reverse engineering, we limited our analysis to a sample of 30 \gps and 30 \rmb malware samples, using the \texttt{SimiDroid}\footnote{https://github.com/lilicoding/SimiDroid},
\texttt{apktool}~\footnote{https://ibotpeaches.github.io/Apktool/},
and \texttt{smali2java}~\footnote{https://github.com/AlexeySoshin/smali2java} tools. Considering this sample, the median \sscore is 0.99 and 0.90 for the \gps and \rmb families, respectively. Table~\ref{tab:simidroid-outputs} and Table~\ref{tab:simidroid-outputs-2} summarize the outputs of \texttt{SimiDroid} for these samples.

Regarding the \gps malware, the similarity assessment of this sample of 30 apps reveals a few modification patterns when comparing the original and the repackaged versions. First, no instance in this \gps sample dataset modifies the Android Manifest file to require additional permissions. In most cases, the repackaged version just changes the Manifest file to modify either the package name or the main activity name. Moreover, 29 out of the 30 samples in this dataset  {\bf modifies} the 
method \texttt{void onReceive(Context, Intent)} of the class \texttt{com.games.AdReciver}. Although the results of the decompilation process are difficult to understand in full (due to code obfuscation), the goal of this modification is to change the behavior of the benign version, so that it can download a different version of the \texttt{data.apk} 
asset. Figure~\ref{code:onReceive} shows the code pattern of the \texttt{onReceive} method present in the samples. This modification typically uses a new method (\texttt{public void a(Context)}) in the repackaged versions, often introduced into the same class (\texttt{AdReceiver}). Since there are no additional calls to sensitive APIs, the \mas fails to correctly label the \gps samples. This limitation holds regardless of our experimental choices, such as using the DroidBot tool (instead of more recent test case generation tools) or running the samples for three minutes.

\begin{figure*}
\begin{lstlisting}[language=Java,
    basicstyle=\footnotesize]
public void onReceive(Context context, Intent intent) {
  SharedPreferences sp = context.getSharedPreferences(String.valueOf("com.") + "game." + "param", 0);
  int i = sp.getInt("sn", 0) + 1;
  System.out.println("sn: " + i);
  if (i < 2) {
    mo4a(context);
    SharedPreferences.Editor edit = sp.edit();
    edit.putInt("sn", i);
    edit.commit();
  } else if (!new C0004b(context).f7h.equals("")) {
    String str1 = context.getApplicationInfo().dataDir;
    String str2 = String.valueOf(str1) + "/fi" + "les/d" + "ata.a" + "pk";
    String str3 = String.valueOf(str1) + "/files";
    String str4 = String.valueOf("com.") + "ccx." + "xm." + "SDKS" + "tart";
    String str5 = String.valueOf("InitS") + "tart";
    String str6 = "ff048a5de4cc5eabec4a209293513b6e";    
    C0003a.m3a(context, str2, str3, str4, str5, str6);
    SharedPreferences.Editor edit2 = sp.edit();
    edit2.putInt("sn", 0);
    edit2.commit();
  }
}
\end{lstlisting}
\caption{Method introduced in 29 out of 30 \gps malware we randomly selected from the \cds.}
\label{code:onReceive}
\end{figure*}

Our assessment also reveals recurrent modification patterns that {\bf delete} methods in the
repackaged version of the apps. For instance, 20 repackaged apps in our
\gps sample of 30 malware remove the method \texttt{void b(Context)} from the
class \texttt{com.game.a}. This class extensively uses 
the Android reflection API. Although it is not clear the real purpose of removing these methods,
that decision simplifies the procedure of downloading a \texttt{data.apk} asset that is different
from the asset available in the original version of the apps. Removing those methods might
also be a strategy for antivirus evasion. For instance, although
some usages of the class \texttt{DexClassLoader} might be legitimate, it allows specific
types of attack based on dynamic code injection~\cite{DBLP:conf/acsac/FalsinaFZKVM15}. As such, 
antivirus might flag specific patterns using the Android reflection API suspect. 
Unfortunately, the \mas also fails to identify
a malicious behavior with this type of change (i.e., changes that remove methods), again, regardless of the decisions we follow in our experiment.
Listing~\ref{code:deletedMethod} shows an example of code pattern frequently removed
from the repackaged versions from the \gps family. 

\begin{figure*}[t]

\begin{lstlisting}[language=Java,
    basicstyle=\footnotesize]
public static void m7a(Activity activity, String str, String str2, String str3, String str4, String str5) {
  try {
    Class loadClass = new DexClassLoader(str, str2, (String) null, activity.getClassLoader()).loadClass(str3);
    Object newInstance = loadClass.getConstructor(new Class[0]).newInstance(new Object[0]);
    Method method = loadClass.getMethod(str4, new Class[]{Activity.class, String.class});
    method.setAccessible(true);
    method.invoke(newInstance, new Object[]{activity, str5});
  } catch (Exception e) {
    e.printStackTrace();
  }
}  
\end{lstlisting}
\caption{Example of method that is typically removed from the repackaged apps of the \gps family.}
\label{code:deletedMethod}
\end{figure*}

\begin{table}[ht]
  \centering
  \caption{Summary of the outputs of the \texttt{SimiDroid} tool for the sample of 30
    \gps malware. (IM) Identical Methods, (SM) Similar Methods, (NM) New Methods, and
    (DM) Deleted Methods.}
  \begin{tabular}{lrrrrrr}
    \hline
 Hash & \sscore &   IM  &   SM  &  NM   &  DM  \\ \hline
 33896E & 0.9994 &  3205 &     2 &     0 &     0 \\ 
 0C962D  & 0.9994 &  3413 &     1 &     1 &    10 \\ 
 BCDF91  & 0.9992 &  2645 &     2 &     0 &     0 \\ 
 01ECE4  & 0.9991 &  5697 &     4 &     1 &    10 \\ 
 A306DA  & 0.9989 &  1886 &     1 &     1 &     6 \\
 4010CA  & 0.9987 &  3721 &     1 &     4 &     6 \\
 5B5F2D  & 0.9983 &  1164 &     2 &     3 &     0 \\
 010C07  & 0.9982 &  2248 &     4 &     3 &     0 \\
 F9FC04  & 0.9982 &  1121 &     1 &     1 &     6 \\
 E29F53  & 0.9976 &   842 &     1 &     1 &     6 \\
 FE76EB  & 0.9976 &   839 &     1 &     1 &     6 \\
 842BD5  & 0.9973 &  2249 &     3 &     3 &     3 \\
 295B66  & 0.9972 &  1081 &     2 &     1 &    10 \\
 92209D  & 0.9971 &   698 &     2 &     3 &     0 \\
 0977B0  & 0.9969 &  1613 &     4 &     1 &    10 \\
 347FCF  & 0.9967 &   613 &     1 &     1 &     6 \\
 00405B  & 0.9965 &   864 &     2 &     1 &    10 \\
 67310E  & 0.9957 &  1164 &     2 &     3 &     3 \\
 CCD29E  & 0.9954 &   436 &     2 &     0 &     0 \\
 610113  & 0.9941 &   836 &     4 &     1 &    10 \\
 A871E0  & 0.9941 &   836 &     4 &     1 &    10 \\
 ECEA10  & 0.9913 &   229 &     1 &     1 &     6 \\
 E53FAA  & 0.9889 &   267 &     2 &     1 &    10 \\
 723C23  & 0.9870 &   228 &     2 &     1 &    10 \\
 D95B6E  & 0.9870 &   833 &    10 &     1 &    10 \\
 17722D  & 0.9743 &   265 &     6 &     1 &    10 \\
 537492  & 0.9504 &   134 &     6 &     1 &    10 \\
 078E0A  & 0.9504 &   134 &     6 &     1 &    10 \\
 D83F1C  & 0.9494 &   150 &     2 &     6 &     6 \\
 E5D716  & 0.8840 &  2035 &    68 &   199 &   199 \\

   \hline
 \end{tabular}
 \label{tab:simidroid-outputs}
\end{table}

 \begin{table}[ht]
  \centering
  \caption{Summary of the outputs of the \texttt{SimiDroid} tool for the sample of 30
    \rmb malware. (IM) Identical Methods, (SM) Similar Methods, (NM) New Methods, and
    (DM) Deleted Methods. }
  \begin{tabular}{lrrrrrr}
    \hline
 Hash & \sscore &   IM  &   SM  &  NM   &  DM  \\ \hline

14BBE2  &	0.9940 &3348&6	&532	&14\\
BFEF74  &	0.9940 &3348&6	&532	&14\\
A3FACA  &	0.7918 &2667&80&	112&1	621\\
10F22D  &	0.9940 &3348&6	&532	&14\\
50193A  &	0.9940 &3348&6	&532	&14\\
5A7536  &	0.9940 &3348&6	&532	&14\\
BCC0DB  &	0.7918 &2667&80&	112&1	621\\
E866CB  &	0.9940 &3348&6	&532	&14\\
CDD316  &	0.9940 &3348&6	&532	&14\\
DF39F6  &	0.7918 &2667&80&	112&1	621\\
3FFAFF  &	0.9121 &3072&184&	628&	112\\
C8C63D  &	0.9940 &3348&6	&532	&14\\
48C562  &	0.9121 &3072&184	&628&	112\\
D27F26  &	0.7918 &2667&80&	112&1	621\\
F4BBEC  &	0.9121 &3072&184	&628&	112\\
BCF14C  &	0.9127 &3074&182	&628&	112\\
7FBF11  &	0.7918 &2667&80&	112&1	621\\
9D35D4  &	0.7918 &2667&80&	112&1	621\\
D1B27E  &	0.9940 &3348&6	&532	&14\\
94DD4B  &	0.9940 &3348&6	&532	&14\\
2D217E  &	0.7918 &2667&80&	1121&	621\\
66F167  &	0.7918 &2667&80&	1121&	621\\
155D4A  &	0.9940	&3348	&6	&532	&14\\
8CB780  &	0.9127	&3074	&82	&628	&112\\
C251FA  &	0.9940	&3348	&6	&532	&14\\
40487B  &	0.7918	&2667	&80	&1121	&621\\
F29692  &	0.9940	&3348	&6	&532	&14\\
0E3679  &	0.9127	&3074	&182	&628	&112\\
7A4F31  &	0.9121	&3072	&184	&628	&112\\
BB3EDE  &	0.7105	&2393	&256	&1217	&719\\

   \hline
 \end{tabular}
 \label{tab:simidroid-outputs-2}
\end{table}

In summary, our reverse engineering effort brings evidence that malware samples from the \gps family neither modify the Android Manifest files nor call additional sensitive APIs. It acts as a
downloader for further malicious app~\cite{DBLP:conf/ndss/ArpSHGR14}---which reduces the ability of the \mas to classify a sample as a malware correctly. Both versions (original/repackaged) from the \gps family have the same behavior for showing advertisements to the user,
however, the repackaged version has additional call sites to the advertisement API and the advertisement sources are different. 

Similar to the approach used for \gps samples, we also reverse-engineered a random selection of 30 samples from the \rmb family that were not detected by the \mas. As with the \gps family samples, no instance from the \rmb family modifies the Manifest file or inserts extra calls to sensitive APIs, making it harder for the \mas to label the samples as malware correctly. However, our reverse engineering reveals that all apps store a file with the extension ``.so'' (Shared Object files) in their lib directory. These files are dynamic libraries containing native code written in C or C++, and are often used by apps for performance reasons, when resource-intensive tasks need to be performed~\cite{ruggia2023dark}.

Unfortunately, creating malicious repackaged apps using ``.so'' files is also possible, as they can be replaced by a version containing harmful code~\cite{DBLP:conf/dsn/QianLSC14}. The Shared Object files also allow for attacks based on dynamic code injection~\cite{DBLP:conf/acsac/FalsinaFZKVM15}, considering that Android apps can use methods like \texttt{System.loadLibrary()} or \texttt{System.load()} to download malicious ``.so''  files from a remote server. Once on the device, malicious apps can use these files to interface with Java code in Android apps, via the Java Native Interface (JNI), performing malicious operations on low-level code and bypassing security mechanisms, like the \mas.

Our assessment confirms that all \rmb samples contain Java code that loads a native library. In particular, to load these libraries, the samples use the \texttt{loadLibrary} method of the \texttt{System} class, which is called in the static constructor of the \texttt{mainActivity} class. The \texttt{loadLibrary} method takes ``game'' as an argument, and the code automatically searches the default lib directory for the \texttt{.so} file named (\texttt{lib}+\texttt{argument}). The ``lib'' directory contains the \texttt{libgame.so} file in all samples. Figure~\ref{code:jni} presents the code pattern of the \texttt{mainActivity} class found in the samples from \rmb family.

\begin{figure}[ht]
\begin{lstlisting}[language=Java,
    basicstyle=\footnotesize]

public class PZPlayer extends Cocos2dxActivity {
    // ...
    System.loadLibrary("game");
    // ...
}
\end{lstlisting}
\caption{Thie code links this java file into libgame shared library}
\label{code:jni}
\end{figure}

The \texttt{libgame.so} file contains compiled code written in C or C++, which is loaded into memory and linked to the apps at runtime. Although machine code is difficult to analyze, all the files include the \texttt{JNI\underline{\hspace{.1in}}OnLoad} function, which the JNI implementation automatically uses to link Java methods and native functions. When we analyzed the ``.so'' file, we found that they all differ in size and content between the original and repackaged apps. It is possible that changes of interest occurred in the native \texttt{libgame.so} file and have gone unnoticed by the \mas. Again, since the \mas only considers differences in calls to sensitive APIs, it is unlikely to correctly classify these samples using other test case generation tools or by extending the execution time during its exploratory phase.

\section{Discussion}\label{sec:discussion}

In this section, we answer our research questions,
summarize the implications of our results, and discuss possible
limitations of our study that might threaten the
validity of the results presented so far.

\subsection{Answers to the Research Questions}

The results we presented in the previous sections
allow us to answer our three research questions, as
we summarize in the following.

\begin{itemize}
\item \textbf{Performance of the \mas (RQ1).} 
Our study indicates that the accuracy of the \mas reported in previous studies~\cite{DBLP:conf/wcre/BaoLL18,DBLP:journals/jss/CostaMMSSBNR22} does not generalize to a larger dataset. That is, while in our reproduction study (using the \sds of previous research) the \mas leads to an accuracy of \fscoreSmall, we observed a drop of precision and recall that leads to an accuracy of \fscore in the presence of our \cds (\apps pairs of original and repackaged versions of Android apps). 

\item \textbf{Similarity Analysis (RQ2).} Our results bring evidence about the association between the similarity of the original and repackaged versions of an app and the ability of the \mas to correctly classify a repackaged version of an app as a malware. Therefore, the similarity assessment is relevant for explaining the performance of the \mas to classify certain repackaged versions of an app as malware.

\item \textbf{Malware Family Analysis (RQ3).} The results indicate that some families are responsible for the largest number of false negatives in the complete dataset. We specifically further investigate the \gps and \rmb families---a particular type of Adware, designed to display advertisements while an app is running automatically. After reverse engineering a sample of 60 malware apps from \gps and \rmb family, we confirmed that the \mas cannot identify the patterns of changes introduced in the repackaged versions of the apps. The prevalence of the \gps and \rmb families in the Android malware landscape accounts for the poor performance of the \mas in malware classification on the large dataset.

\end{itemize}

\subsection{Implications}\label{sec:implications} 

Contrasting to previous research works~\cite{DBLP:conf/wcre/BaoLL18,DBLP:conf/iceccs/LeB0GL18,DBLP:journals/jss/CostaMMSSBNR22},
our results lead to a more systematic understanding
of the strengths and limitations of using the \mas
for malware classification. In particular, this is the
first study that empirically evaluates the \mas
considering as ground truth the outcomes
of \vt---a common practice in the malware identification research. This decision allowed us to explore the
\mas performance using well-known accuracy metrics (i.e., precision, recall, and
$F_1$ score). Contrasting with previous studies that assume that all repackaged versions of the
apps were malware. Our triangulation with \vt reveals this is not true. Although
the \mas presents a good accuracy for the \sds ($F_1$ = \fscoreSmall), 
in the presence of a large dataset the \mas accuracy drops significantly ($F_1$ = \fscore). 

We also reveal that some families in the \cds are responsible for a large number of false negatives,
compromising the accuracy of the \mas.
Altogether, the takeaways of this research are twofold:

\begin{itemize}
  \item Negative result: the \mas for malware detection exhibits a much higher false negative rate than previous research reported. 

  \item Future directions: Researchers should advance the \mas for malware detection by exploring more sophisticated techniques to differentiate between benign and malicious apps. In particular, since our reverse engineering results suggest that \gps and \rmb---two recurrent malware families---use the network to download new assets, new approaches might benefit from monitoring not only calls to sensitive APIs but also network traffic, as well as mining sensitive calls to native APIs embedded in \texttt{so} files. The versatility of the Java Native Interface (JNI) has introduced challenges. Malware authors increasingly use the native layer to hide malicious code, making both static and dynamic analysis more difficult. The current state of the art in sandbox mining overlooks native calls.

\end{itemize}

\subsection{Threats to Validity}\label{sec:threats}


There are some threats to the validity of our results.
Regarding {\bf external validity}, one concern relates to the 
representativeness of our malware datasets and how generic our findings are.
Indeed, mitigating this threat was one of the motivations for our research,
since, in the existing literature on the \mas for malware classification, researchers had explored just
one dataset of 102 pairs of original/repackaged apps. Curiously,
for this small dataset, the performance of the
\mas is substantially superior
to its performance on our \cds (\apps pairs of
apps).

We contacted the authors of the Bao et al. original research paper~\cite{DBLP:conf/wcre/BaoLL18}, asking them
if they had used any additional criteria for selecting the pairs of apps in their
dataset. Their answers suggest the contrary: they have not used
any particular app selection process that
could explain the superior performance of the \mas for the \sds. We believe 
our results in the \cds generalize better than previous research work,
since we have a more comprehensive collection of malware with different
families and degrees of similarity. Nonetheless, our
research focuses only on Android repackaged malware. Thus, we cannot generalize
our findings to malware that targets other platforms or uses different approaches
to instantiate a malicious asset. Besides that, repackaging is a recurrent approach
for implementing Android malware.

Regarding {\bf conclusion validity}, during the exploratory phase of the \mas, we collected the set of calls to sensitive APIs the original version of an app executes, while running a test case generation tool (DroidBot).
In the exploratory phase, the \mas assumes the existence of a benign original
version of a given app. \highlight{We also query \vt to confirm this
assumption, and found that the original version of seven (out 102) apps in the
\sds contains malicious code. We believe the authors of previous studies carefully check that assumption,
and this difference had occurred because the outputs of \vt change over time~\cite{DBLP:conf/uss/ZhuSYQZS020}}, and a dataset that is consistent on a given date may not remain consistent in the future. Therefore, while reproducing this research, it is necessary to query \vt to get the most up-to-date classification of the assets, which might lead to results that might slightly
diverge from what we have reported here. Besides that, in the \cds we only consider
pairs of original/repackaged apps for which \vt classifies the original version as benign. 

Regarding {\bf construct validity}, we address the main threats to our study by using simple and
well-defined metrics that are in use for this type of research: number of malware samples the
\mas correctly/wrongly classify in a dataset (true positives/false negatives).
We computed the accuracy results using precision and recall based on these metrics. In a preliminary study, we
investigated whether or not the \mas would classify an original version of an app as malware,
computing the results of the test case generation tools in multiple runs. After combining three executions
in an original version to build a sandbox, we did not find any other execution that could wrongly
label an original app as malware. Also, we label a repackaged version of an app as malware
  only if \vt reports that at least two engines detect suspicious behavior in that asset.
  This decision might be viewed as either a weak or strong constraint and could
  raise concerns about construct validity. However, when we relax this constraint and label an asset as malware whenever at least one engine detects suspicious behavior, precision improves to 0.85,
  but recall drops to 0.39. Overall, the accuracy of the \mas remains almost unchanged (F$_1$ = 0.53)---still significantly lower than the precision of the \mas for \sds.
  We also evaluated accuracy by considering an asset as malware when at least five or ten \vt security engines flagged it. As shown in Table~\ref{tab:engines},
  the results did not diverge significantly from what we have reported in this paper.

\begin{table*}[htb]
  \caption{Accuracy of the \mas at \cds (4,076 pairs) based on engines.}
\centering{
  \begin{tabular}{lrrrrrr} \hline
    Engine(s) & TP   & FP  & FN  & Precision & Recall & $F_1$ \\
    \hline
    At least 01    & 1,222  & 220 & 1,900 & 0.85       & 0.39   & 0.53  \\
   
    At least 02    & 1,175  & 220 & 1,720 & 0.84       & 0.40   & 0.54  \\
   
    At least 05    & 1,087  & 220 & 1,578 & 0.83       & 0.40   & 0.54  \\
   
    At least 10    & 1,002  & 220 & 1,469 & 0.81       & 0.40   & 0.54  \\
    \hline
    
  \end{tabular}
  }
  \label{tab:engines}
\end{table*}

\section{Conclusions}\label{sec:conclusions}

To better understand the strengths and limitations of the \mas for repackaged malware detection, this paper reported the results of an empirical study that replicates previous research works~\cite{DBLP:conf/wcre/BaoLL18,DBLP:journals/jss/CostaMMSSBNR22}. The study utilizes a more diverse dataset compared to those used in previous research, with the aim of providing a more comprehensive evaluation of the approach. To our surprise, compared to published results, the performance of the \mas drops significantly for our comprehensive dataset ($F_1$ score reduces from \fscoreSmall in previous papers to 0.54 here). This result is partially explained by the high prevalence of specific malware families (named \gps and \rmb), whose samples are incorrectly classified by the \mas. We also report the results of a reverse engineering effort, whose goal was to understand the characteristics of the \gps and \rmb family that reduce the performance of the \mas for malware classification. Our reverse engineering effort revealed common changing patterns in the \gps repackaged versions of original apps, which mostly use reflection to download an external apk asset for handling advertisements without introducing additional calls to sensitive APIs. Similarly, the \rmb family does not include any additional calls to sensitive resources; however, it often uses JNI to interact with native code, which can be used to perform malicious operations at a low level, compromising the effectiveness of the \mas for malware identification. These negative results highlight the current limitations of the \mas for malware classification and suggest the need for further research to integrate the \mas with other techniques for more effective malware identification.

\balance 

\bibliographystyle{IEEEtran}
\bibliography{ref}

\end{document}